\documentclass[12pt]{article}

\usepackage{amssymb,latexsym}

\makeatletter \@addtoreset{equation}{section} \makeatother
\renewcommand{\theequation}{\thesection.\arabic{equation}}

\newcounter{parentequation}
\newenvironment{subequations}{%
  \refstepcounter{equation}%
  \begingroup
\let\protect\noexpand
  \edef\@tempa{\def\noexpand\theparentequation{\theequation}}%
  \expandafter
  \endgroup\@tempa
  \setcounter{parentequation}{\value{equation}}%
  \setcounter{equation}{0}%
  \def\theequation{\theparentequation\alph{equation}}%
  \ignorespaces
}{%
  \setcounter{equation}{\value{parentequation}}%
}

\begin{document}

\begin{titlepage}

\thispagestyle{empty}

\begin{flushright}
\hfill{CERN-TH/2005-021} \\
\hfill{hep-th/0502066}
\end{flushright}

\vspace{35pt}

\begin{center}{ \LARGE{\bf 
Gauged supergravity algebras from \\[2mm]
twisted tori compactifications with fluxes.
}}

\vspace{60pt}

{\bf  Gianguido Dall'Agata \  and \ Sergio Ferrara}

\vspace{15pt}

{\it  
Physics Department,\\
Theory Unit, CERN, \\
Geneva 23, CH1211,
Switzerland}

\vspace{70pt}

{ABSTRACT}

\end{center}

\vspace{20pt}
 
Using the equivalence between Scherk--Schwarz reductions and twisted
tori compactifications, we discuss the effective theories obtained by
this procedure from M--theory and ${\cal N} =4$ type II orientifold
constructions with Neveu--Schwarz and Ramond--Ramond form fluxes
turned on.
We derive the gauge algebras of the effective theories describing
their general features, in particular the symplectic embedding in the 
duality symmetries of the theory.
The generic gauge theory is non--abelian and its gauge group is given
by the semi-direct product of subgroups of $SL(7)$ or $SL(p-3)\times
SL(9-p)$ for $p=3,\ldots,9$, with generators describing nilpotent
subalgebras of $e_{7(7)}$ or $so(6,6)$ (in M and type II theories
respectively).

\end{titlepage}

\newpage

\baselineskip 6 mm

\section{Introduction}

The addition of fluxes to string theory backgrounds is an important
perturbative mechanism to address the moduli stabilization problem.
When considering the effective theory, the flux backreaction on the
geometry is often neglected and the potential is expressed in terms of
the fluxes and the geometrical structures of the original manifold.
For instance, in type IIB ${\cal N} = 1$ compactifications on
Calabi--Yau manifolds $Y_6$, the addition of fluxes induces a
superpotential on the effective theory which reads \cite{Gukov:1999gr}
\begin{equation}
W= \int_{Y_6} G \wedge \Omega.
\label{pot1}
\end{equation}
This expression depends on the complex 3--form flux $G$ and on the
holomorphic form on the Calabi--Yau.
This allows to fix the axio/dilaton moduli as well as all the
complex--structure deformations.
This type of approximation is usually justified by the fact that the
introduction of fluxes often leads simply to the addition of a
conformal factor in front of the internal manifold.
When this happens, at least in the ``small flux'' approximation, one
is allowed to describe the moduli of the effective theory in terms of
the moduli of the original flux-less background.

Generally, however, the fluxes backreaction deforms non--trivially
the geometry of the internal manifold and this deformation must also
induce a change to the effective potential.
For the Heterotic string it has been shown
\cite{Cardoso:2002hd,Becker:2003sh} that
such modifications are captured by the following superpotential:
\begin{equation}
W = \int \left(H + i \, dJ\right) \wedge \Omega.
\label{pot2}
\end{equation}
This expression takes into account the appearance of a twist
on the internal manifold which reflects in $dJ \neq 0$.
Since also the K\"ahler form appears in the potential, there is
the possibility that all the moduli may be fixed.
Obviously the generic moduli space of such internal manifolds can be
very different from that of Calabi--Yaus and it is not clear how to
discuss the effective scalar $\sigma$--model for such
compactifications.
There is however at least one case where the effective theory should
have an easy description: the twisted tori solutions.

It is known since the original papers by Scherk and Schwarz
\cite{Scherk:1979zr} that their mechanism to introduce masses in
effective theories coming from string compactifications on tori is
equivalent to compactifying the same theories on twisted versions of
the same manifolds, where one deforms both the complex and K\"ahler
structures.
The twist can be read in terms of the original moduli of the straight
torus as the introduction of mass couplings.
In this way one has some control on both the complex and K\"ahler
deformations of the internal manifold from the effective theory point
of view.

Such an idea was recently used in \cite{Derendinger:2004jn}, following
\cite{Kaloper:1999yr}, to show that in type IIA compactifications on
twisted $T^{6}/Z_2\times Z_{2}$ all the moduli can be stabilized.
It is natural to try to extend this type of analysis to more general
compactifications in M--theory as well as in type IIA and IIB
supergravity.
One should first determine which are the deformations of the effective
gauged supergravities due to the twisting of the internal manifold (in
addition to the fluxes).
Then one should compute the scalar potentials and analyze moduli
stabilization.
Finally, one should interpret the values of the moduli at the fixed
points in terms of the geometry of the internal manifolds, possibly
giving the expressions for $dJ$ and $d\Omega$.
Part of this analysis (though in a complementary approach) was carried
out for some type II compactifications on twisted tori which result
from series of T--dualities on flat $T^{6}/Z_2$ in
\cite{Kachru:2002sk,Schulz:2004ub}.
There the authors argued the form of the effective superpotential and
determined the conditions to obtain supersymmetric vacua.

Here we will focus on M--theory compactifications on a twisted $T^7$
with a physical 4--form flux and on the ${\cal N} = 4$ type II
orientifold compactifications presented in
\cite{Angelantonj:2003rq,Angelantonj:2003up} to which we add a twist 
of the internal geometry.
In particular, we will complete the first step of the analysis
detailing the structure of the gauge algebras of the effective
theories and their embedding in the electric group, marking the
differences with respect to
\cite{Angelantonj:2003rq,Angelantonj:2003up}.
The resulting general structure of the algebra can be summarized by
the commutators of two type of generators: Kaluza--Klein generators
$Z_A$ and tensor gauge generators $X_\alpha$.
The first type of generators gives the symmetries of the internal
metric, while the second one gives the gauge symmetries of the
10--dimensional form fields.
The structure that we obtain is the following:
\begin{equation}
\begin{array}{rcl}
[X_\alpha,X_\beta] &=& 0,\\[2mm] 
[Z_A,X_\alpha] &=&
\left(f_{A\alpha}{}^{\beta} + \tau_{A\alpha}{}^\beta
\right)X_{\beta},\\[2mm] 
[Z_A,Z_B] &=& f_{AB}{}^\alpha X_\alpha + \tau_{AB}{}^C
Z_{C}.\\[2mm]
\end{array}
\label{genalgebra}
\end{equation}
The non--trivial structure constants are related to either the
appearance of fluxes $f$, or to the Scherk--Schwarz reduction $\tau$.
These structure constants must satisfy some non--trivial constraints
(following also from the Jacobi identities) which can be interpreted
in terms of the original 10 or 11--dimensional theory as the Bianchi
identities of the fluxes or as the invariance of the volume measure of
the internal manifold under change of coordinates.
We point out that the Lie algebra structure in (\ref{genalgebra})
strictly applies only when additional relations among $\tau$ and $f$
exist, which are stronger than the integrability conditions in $D=10$
or $D=11$.
However, when these latter are satisfied, one can show that
(\ref{genalgebra}) has a non trivial extension to a free differential
algebra, whose ``structure constants'' $f$ for the quadratic part of
the curvature of the 1--form potentials do not fulfill the
relations of an ordinary Lie algebra but rather of a free differential
algebra \cite{inprep}.

The deformation of the effective theory due to the 10-- or
11--dimensional fluxes is equivalent to the gauging of some
Peccei--Quinn isometries of the scalar manifold whose generators form
a nilpotent subalgebra.
The Scherk--Schwarz reduction on the other hand has an action on the
gauge vectors which can be interpreted as that of generators forming a
triangular subgroup of the duality group $E_{7(7)}$ or $SO(6,6)$.
Although this includes a nilpotent part, it does not coincide nor
necessarily contain the one generated from the fluxes.

The general action of the full duality group mixing vector
field--strengths with their duals ($E_{7(7)}\in Sp(56)$ and $SO(6,6)
\times SL(2,{\mathbb R}) \in Sp(24)$ for M--theory and type II
compactifications respectively) is always lower triangular for
Lagrangian theories:
\begin{equation}
S = \left( 
\begin{array}{cc}
a&0\\
c& -a^{T}
\end{array}
\right).
\label{Sduality}
\end{equation}
From the general structure of the algebra (\ref{genalgebra}) we can
only deduce directly the form of the matrix $a$, which is the one
acting on the gauge vectors and may lead to non--abelian gauge groups.
The matrix $c$ on the other hand is determined by the Lagrangian
couplings linear in the axion fields and containing two gauge field
strengths.
This latter has always a nilpotent structure and is obviously fixed by
the gauging procedure through the determination of the Lagrangian.
Since here we are going to focus only on the derivation of the gauge
algebras, we will display the explicit form of $S$ only for 3 simple
examples.
It should be noted that this action of the gauge group on the vector
fields may be there also for abelian theories, where $a=0$.
The interesting new information that we gain from our analysis is that
also the matrix $a$ has a lower triangular form according to the
decomposition of the duality group $E_{7(7)}$ or $SO(6,6)$ with
respect to $SL(7)$ or $SL(p-3) \times SL(9-p)$ and, as said above, it
will include a rotational part as well as a nilpotent one.

A special case of this general analysis is given by the
Scherk--Schwarz flat groups described in the context of ${\cal N} =8$
supergravity \cite{Cremmer:1979uq,Andrianopoli:2002aq}.
Indeed the structure of those groups turns out to be a particular case
of the ones considered here with respect to the structure of the $a$
and $c$ matrices shown above.

This paper is organized as follows: in section \ref{Mthsec} we 
consider M--theory on a twisted 7--torus with 4--form flux turned on 
and show how the M--theory gauge algebra is embedded in a lower 
triangular symplectic representation of $E_{7(7)}$.
In section \ref{orientsec} the analysis is extended to the type II 
orientifolds with fluxes on twisted 6--tori.
Here the Scherk--Schwarz generators are those of $SL(p-3)$ and the 
embedding of the gauge algebra is presented as a lower--triangular 
representation of $SO(6,6)$, due to the $SO(6,6)$ grading with 
respect to $SO(1,1) \times SO(1,1) \times SL(p-3) \times SL(9-p)$ 
(the second $SO(1,1)$ being absent for $p=3,9$).
We finish in section \ref{finalsec} with some comments.

\section{M--theory on twisted ${\mathbb T}^7$ with fluxes}
\label{Mthsec}

The strategy we are going to employ in the following is to determine
the gauge algebra of the effective theories by looking at the gauge
and Kaluza--Klein transformations of the metric and form fields which,
once reduced on the internal manifold, give vector fields in 4d.
As the gauge algebra must close, so must the products of gauge
transformations on the vector fields.
Moreover, since vector fields span a faithful representation of the
gauge group, we can derive {\it all} the commutators by evaluating the
product of couples of infinitesimal transformations.
Details of such a method can also be found in \cite{Kaloper:1999yr},
where the gauge algebra of the Scherk--Schwarz reduction of Heterotic
theory in the presence of fluxes was obtained.

As a simple and instructive example which captures most of the general
features of Scherk--Schwarz reductions in the presence of fluxes we
now focus on M--theory compactifications on twisted ${\mathbb T}^7$
with fluxes.
The bosonic field content is simply given by the 11--dimensional
metric $G$ and a 3--form potential ${\cal A}$.
Once these fields are reduced to 4 dimensions one obtains the
4--dimensional metric $G_{\mu\nu}$, 28 vector fields, 7 coming from
the metric $G_{\mu M}$ and 21 from the 3--form ${\cal A}_{\mu MN}$,
63 scalars, 28 coming from the metric $G_{MN}$ and 35 from the
3--form ${\cal A}_{MNP}$ and 7 antisymmetric tensors ${\cal 
A}_{\mu\nu M}$, which, under certain conditions, can also be dualized 
to 7 scalars $A^{M}$, for a total of 70 scalars.

In the reduction process the metric vectors will always maintain the
same transformation rules regardless of the theory we start from.
These are vectors coming from the reduction of the 11-- (or
10--)dimensional metric tensor with one space--time index and one
index on the internal space, for this reason adding fluxes to the form
fields will never change their gauge transformation rule which reads
\cite{Scherk:1979zr}
\begin{equation}
\delta G_{\mu}^M = \partial_\mu \omega^M - \tau_{NP}^M \omega^N
G^P_\mu.
\label{delG}
\end{equation}
This gauge transformation is a consequence of the invariance of the
original theory under diffeomorphisms of the internal 
manifold\footnote{A paper has appeared \cite{Andrianopoli:2005jv} where the equivalence 
between general Scherk--Schwarz phases and toroidal backgrounds with 
torsion and fluxes has been proved.}.

As noted by Scherk and Schwarz and as more precisely stated by
Kaloper and Myers in \cite{Kaloper:1999yr}, the Scherk--Schwarz
reduction is essentially equivalent to a compactification on twisted
tori.
This means that if one labels the vielbeins of the internal manifold
as $e^M$, the structure constants $\tau$ appearing in (\ref{delG}) are
related to the non--trivial connection of the internal manifold
\begin{equation}
d e^M = \frac12 \tau^M_{NP} e^{N}\wedge e^P,
\label{tors}
\end{equation}
for constant coefficients $\tau$.
From this fact one also derives the main constraints that the
structure constants must obey and which are also needed in order to
fulfill the Jacobi identities of the resulting gauge algebra in 4
dimensions.
Closure of the differential on the vielbeins $d^2 = 0$ applied to
(\ref{tors}) gives
\begin{equation}
\tau_{[NP}^{S}\tau_{Q]S}^M = 0.
\label{const1}
\end{equation}
The other condition,
\begin{equation}
\tau_{MN}^N = 0,
\label{tracetau}
\end{equation}
comes instead from the requirement that the reduced action be 
invariant against internal coordinate transformations 
\cite{Scherk:1979zr} and has also the interpretation of invariance of 
the measure of the internal volume under the same transformations 
\cite{Kaloper:1999yr}.

In this reduction, the other 4--dimensional vector fields come from
the reduction of the 11--dimensional 3--form ${\cal A}$.
We will now detail the derivation of the transformation rules for such
vector fields in order to explain the strategy which we are also going
to use for the reduction of all the other forms in the type II
examples, though not giving all the details as for this one.

All the fields coming from the reduction of the 3--form
field ${\cal A}$ are subject to transformations under the relic of the
11--dimensional gauge symmetry ${\cal A} \to {\cal A} + d\Sigma$ as
well as of the diffeomorphisms ${\cal A} \to {\cal A} + {\cal
L}_{\omega} {\cal A}$.
Since the Lie derivative of a scalar form field can be expressed as
${\cal L}_{\omega} \equiv i_{\omega}d + d i_{\omega}$, we can see that
the general variation of ${\cal A}$ under these combined
transformations is
\begin{equation}
\delta {\cal A} = i_{\omega} d{\cal A} + d i_{\omega} {\cal A} + d\Sigma.
\label{a0}
\end{equation}
This action is then inherited by the reduced degrees of freedom which
come from the decomposition of ${\cal A}$ in external and internal
components.
From the 11--dimensional 3--form one gets 7 4--dimensional vector
fields, 28 scalars and a cosmological constant term according to the
decomposition
\begin{equation}
\begin{array}{rcl}
{\cal A}(x,y) &=& \displaystyle \frac{1}{3!} \, {\cal A}_{\mu\nu\rho} dx^\mu \wedge 
dx^\nu \wedge dx^\rho + 
\frac{1}{2} \,{\cal A}_{\mu\nu M}(x) dx^\mu \wedge 
dx^\nu \wedge e^{M}(y) \\[3mm]
&+& \displaystyle  \frac12\,
{\cal A}_{\mu MN}(x) dx^\mu \wedge e^M(y) \wedge e^N(y) 
+ \frac{1}{3!} \,{\cal A}_{MNP}(x) 
e^M(y) \wedge e^{N}(y) \wedge e^P(y),
\end{array}
\label{decoA}
\end{equation}
where $e^M$ are the internal manifold vielbeins which reduce to $e^M =
dy^M$ for flat torus compactifications.
When fluxes are introduced (\ref{decoA}) must change, but in a way
that the correct transformation rules for the reduced fields are
obtained.
An appropriate ansatz for such a reduction is given by simply adding
to the right hand side of (\ref{decoA}) a 3--form $\sigma = \sigma(y)$
whose field--strength gives the flux
\begin{equation}
{\cal A} = a + \sigma,
\label{deccA}
\end{equation}
where $a$ contains the expansion (\ref{decoA}) and $d\sigma = g =
\frac{1}{4!} \,g_{MNPQ} \, e^M \wedge e^N \wedge e^P \wedge e^Q$.
From this we can read the transformation on the 4--dimensional fields
\begin{equation}
\delta a = i_{\omega} da + d i_{\omega} a + d \Sigma + i_{\omega} g.
\label{varA}
\end{equation}
Finally, in order to get 4--dimensional fields which transform 
covariantly with respect to the gauge transformations, one also 
employs the composite definitions \cite{Scherk:1979zr}
\begin{eqnarray}
A_{MN\mu} & \equiv & {\cal A}_{MN\mu} - {\cal A}_{MNP}G^{P}_{\mu}\,, 
\label{AMNmudef}\\[2mm]
A_{M\mu\nu} & \equiv & {\cal A}_{M\mu\nu}+ {\cal A}_{MN\mu}G^{N}_{\nu}
- {\cal A}_{MN\nu}G^{N}_{\mu}+ {\cal A}_{MNP} G^{N}_{\mu} G^{P}_{\nu}  
, \label{AMmunu}\\[2mm]
A_{\mu\nu\rho} & \equiv & {\cal A}_{\mu\nu\rho}- 3 G_{[\mu}^P {\cal 
A}_{\nu\rho]P}
+3 {\cal A}_{[\mu MN}G^{M}_{\nu}G_{\rho]}^N- {\cal A}_{MNP} 
G^{M}_{\mu} G^{N}_{\nu} G^P_{\rho}.
\label{Amunuro}
\end{eqnarray}
The resulting transformations are 
\begin{subequations}
\begin{eqnarray}
\delta {\cal A}_{MNP} &=& \omega^{Q}g_{QMNP}+3 \omega^{Q} 
\tau_{Q[M}^{S}{\cal A}_{NP]S} - 3 \tau^S_{[MN}\Sigma_{P]S},\\[2mm]
\delta A_{\mu MN} &=& \partial_\mu \Sigma_{MN} + 2 G_\mu^P \tau_{P[M}^S 
\Sigma_{N]S} - 2 \omega^Q \tau_{Q[M}^S A_{\mu N] S} - \omega^R 
g_{RMNQ} G_\mu^Q \nonumber\\[2mm]
&-& \tau_{MN}^S \tilde \Sigma_{\mu S},
\label{vectransf} \\[2mm]
\delta A_{\mu\nu M} &=& 2 \partial_{[\mu}\tilde \Sigma_{\nu]M} - 2 
G^{P}_{[\mu} \tau^Q_{PM} \tilde \Sigma_{\nu] Q}+ \omega^Q \tau^S_{QM} A_{\mu\nu S}+
\omega^Qg_{QMNP}G_{\mu}^N G_{\nu}^P \nonumber\\[2mm]
&-&\Sigma_{MN} F^{N}_{\mu\nu},\\[2mm]
\delta A_{\mu\nu\rho} &=& 3\partial_{[\mu}\tilde \Sigma_{\nu\rho]} - 
3 F_{[\mu\nu}^P \tilde \Sigma_{P\rho]} - \omega^Qg_{QMNP}G_{\mu}^M G_{\nu}^N 
G_{\rho}^P.
\end{eqnarray}
\label{aprime}
\end{subequations}
where the non--abelian field--strength coming from the metric vectors 
is 
\begin{equation}
F^{N}_{\mu\nu} = 2 \partial_{[\mu} G^{N}_{\nu]} + 
\tau^{N}_{PQ}G^{P}_{\mu}G^{Q}_{\nu}\,,
\label{Fnonab}
\end{equation}
and we also introduced the redefined gauge parameters $\tilde
\Sigma_{\mu M} = \Sigma_{\mu M} + \Sigma_{M N }G_\mu^N$ and $\tilde
\Sigma_{\mu \nu} = \Sigma_{\mu \nu} - 2\Sigma_{[\mu N }G_{\nu]}^N +
\Sigma_{MN} G^M_\mu G^N_{\nu}$.
It is interesting to point out that the last term in the
transformation of the vector fields $A_{\mu MN}$ is the result of the
gauge invariance of $A_{\mu\nu M}$ which is a scalar when dualized.
Once dualized, this tensor gauge invariance is lost and therefore the
vector fields are subject only to their gauge transformations.
It should be noted, however, that such dualization is not always easy
when fluxes are turned on and that one has to be very careful in the
interpretation of the effective theory
\cite{Louis:2002ny,DallAgata:2003yr,DAuria:2004yi,Louis:2004xi,DAuria:2004tr,Sommovigo:2005fk}.
Only after such dualization one gets the $E_{7(7)}$ duality group and
this is what we are going to discuss in the following.
We will actually see that this dualization is not consistent unless 
the obstruction (\ref{missing}) vanishes.
We note that although the gauge algebra is not touched by this 
dualization when compactifying directly to 4 dimensions, the results 
may change if one obtains the effective theory after intermediate 
steps to 5 or 6 dimensions where some of the tensor fields can be 
dualized to vectors.

At this point we can determine the gauge algebra of the reduced 
theory by looking at the commutators of the gauge transformations of 
the vector fields presented in (\ref{delG}) and (\ref{vectransf}).
Assigning to the gauge generators the labels
\begin{equation}
T_{A} = \{ Z_M, W^{MN} \},
\label{gen4}
\end{equation}
matching the parameters $\{ \omega^M, \Sigma_{MN} \}$ respectively, we
can compute the commutators $[\delta_{\omega_2},\delta_{\omega_{1}}]$ 
and $[\delta_{\omega},\delta_{\Sigma}]$, which can be interpreted as
the commutators $[Z,Z]$ and $[Z,W]$ of the gauge algebra.

The first gives
\begin{equation}
\begin{array}{rcl}
[\delta_{\omega_2},\delta_{\omega_1}] A_{\mu MN} &=& \partial_{\mu} 
\left( \omega_{2}^{P} \omega_{1}^{Q} g_{MNPQ}\right) + 3 G^{Q}_\mu 
\tau_{[QM}^{S} \left( \omega^{P}_{2} \omega_1^{R} g_{N]PRS}\right) 
\\[4mm]
&-& 2 \left(\tau^Q_{PR} \omega^{P}_{2} \omega_1^{R}\right) 
\tau_{Q[M}^{S} A_{\mu N]S} - \left( \tau^Q_{PR} \omega^{P}_{2} 
\omega_1^{R}\right) g_{QMNS} G_{\mu}^{S},
\end{array}
\label{comm1}
\end{equation}
provided one uses 
\begin{equation}
\tau^{R}_{[MN}g_{NPQ]R} = 0  \quad \hbox{ 
and } \quad \tau_{[MN}^{S} 
\tau_{P]S}^{Q} = 0.
\label{const1bis}
\end{equation}
This commutator can be interpreted as
\begin{equation}
[\delta_{\omega_2},\delta_{\omega_1}] A_{\mu MN} = \delta_{\Sigma} 
A_{\mu MN} + \delta_{\omega} A_{\mu MN},
\label{comp}
\end{equation}
where now $\Sigma_{MN} =  \omega_{2}^{P} \omega_{1}^{Q} g_{MNPQ}$ and 
$\omega^{M} = \tau^M_{PR} \omega^{P}_2 \omega_1^{R}$.
As a result one obtains that
\begin{equation}
\omega_2^P\omega_1^Q [Z_P, Z_Q] =  \tau^M_{PQ} \omega^{P}_2 
\omega_1^{Q} Z_{M} + \omega_{2}^{P} \omega_{1}^{Q} g_{MNPQ} W^{MN},
\label{com1}
\end{equation}
which means
\begin{equation}
[Z_P, Z_Q] =  \tau^M_{PQ} Z_{M} + g_{MNPQ} W^{MN}.
\label{com2}
\end{equation}

The second commutator gives
\begin{equation}
\begin{array}{rcl}
[\delta_{\omega},\delta_{\Sigma}] A_{\mu MN} &=& \partial_{\mu} 
\left( 2 \omega^{P} \tau_{P[M}^{S} \Sigma_{N]S}\right) + 3 G^{Q}_\mu 
\tau_{[QM}^{S} \left( \omega^{P} \tau_{|P|N]}^{T} 
\Sigma_{ST}-\omega^{P} \tau_{PS}^{T} \Sigma_{N]T}\right) 
\\[4mm]
&+& \tau_{MN}^S \left( \partial_{\mu} \omega^{P} \Sigma_{SP} - 
G_{\mu}^{Q} \omega^{P} \tau^{T}_{PQ}\Sigma_{ST}\right)
\end{array}
\label{comm2}
\end{equation}
provided one uses 
\begin{equation}
\tau_{[MN}^{S} \tau_{P]S}^{Q} = 0.
\label{const2}
\end{equation}
This commutator can be interpreted as
\begin{equation}
[\delta_{\omega},\delta_{\Sigma}] A_{\mu MN} = \delta_{\Sigma^{\prime}} 
A_{\mu MN} + \tau_{MN}^{S} \Sigma_{\mu S},
\label{comp2}
\end{equation}
where now $\Sigma^{\prime}_{MN} = 2 \omega^{P} \tau_{P[M}^{S} 
\Sigma_{N]S}$ and $\Sigma_{\mu S} =  \partial_{\mu} \omega^{P} \Sigma_{SP} - 
G_{\mu}^{Q} \omega^{P} \tau^{T}_{PQ}\Sigma_{ST}$.
As a result one obtains that the commutator algebra closes only up to 
a gauge transformation of the tensor field ${\cal A}_{\mu\nu M}$.
Up to this gauge transformation, the algebra of the vector fields 
then contains (\ref{com2}) and
\begin{equation}
\omega^P\Sigma_{RS} [Z_P, W^{RS}] =  2 \omega^{P} \tau_{P[M}^{S} 
\Sigma_{N]S} W^{MN},
\label{combis1}
\end{equation}
which means
\begin{equation}
[Z_P, W^{RS}] =  2\tau^{[R}_{PQ} W^{S]Q}.
\label{combis2}
\end{equation}

Summarizing, the resulting gauge algebra is
\begin{equation}
\begin{array}{rcl}
[Z_M,W^{NP}] &=&  2\tau^{[N}_{MQ} W^{P]Q},\\[2mm]
[Z_M, Z_N] &=& g_{MNPQ} W^{PQ}+ \tau_{MN}^P Z_P,
\end{array}
\label{algebraM}
\end{equation}
where in order to obtain this result one has to impose some
constraints on the structure constants, namely $\tau_{MN}^N = 0$,
$\tau_{[MN}^{S} \tau_{P]S}^Q = 0$ and $\tau_{[MN}^S g_{PQR]S} = 0$.
We have already seen the interpretation of the first two constraints
from the 11--dimensional point of view.
The last one also has an easy interpretation as the Bianchi identity
of the 3--form gauge field ${\cal A}$ when acting with the
differential on the internal space $dg = 0$.
The Jacobi identities of the full algebra (\ref{algebraM}) close when
the flux is zero or when the Scherk--Schwarz torsion is zero, but do
not close if both $g \neq 0$ and $\tau \neq 0$.
Using the constraint (\ref{const1bis}) one gets that the Jacobi $[Z_M,
[Z_N,Z_P]] + cycl.$ closes up to a term of the form
\begin{equation}
W^{QR}\,\tau_{QR}^{S}\, g_{MNPS},
\label{missing}
\end{equation}
which can be compensated by a tensor gauge transformation acting on 
the vector fields.
If one wants to consider collectively all these transformations, then 
one has to make use of a free--differential algebra formalism 
\cite{inprep}.
Note that if (\ref{missing}) vanishes, then (\ref{algebraM}) is an 
ordinary Lie algebra.
This happens when the Scherk--Schwarz reduction is done in directions 
which are orthogonal to the 4--form fluxes, a relation which can have 
non--trivial solutions, as we will see in \ref{othercasessec}.

The gauge invariance in (\ref{aprime}) tells us that the 
antisymmetric tensor $A_{\mu\nu M}$ have a ``magnetic'' coupling to 
the $A_{\mu MN}$ vectors of the type
\begin{equation}
F_{\mu\nu MN} + \tau_{MN}^{P} A_{\mu\nu P}
\label{shift}
\end{equation}
so that $\tau_{MN}^{P}$ can be identified with a ``magnetic'' mass 
term $m_{\Lambda}^{P}$.
From the Chern--Simons term we also have that the ``dual'' 4--form 
flux $\tilde g^{IJK} = \varepsilon^{IJKMNPQ} g_{MNPQ}$ gives an ``electric'' 
contribution to the mass 
\begin{equation}
\tilde g^{IJK} F_{\mu\nu IJ} A_{\rho\sigma K} 
\epsilon^{\mu\nu\rho\sigma},
\label{massele}
\end{equation}
so that $\tilde g^{IJK} = e^{K \Lambda}$.
We see that the consistency condition of \cite{DAuria:2004yi} $e^{M 
\Lambda}m_{\Lambda}^{N} - e^{N \Lambda}m_{\Lambda}^{M} = 0$ becomes
\begin{equation}
\tilde g^{IJ M} \tau_{IJ}^{N} - \tilde g^{IJ N} \tau_{IJ}^{M} = 0\,,
\label{consist}
\end{equation}
which is a consequence of the Jacobi identity 
$\tau_{[IJ}^{Q}g_{MNP]Q} = 0$ of our gauge algebra.
This is not surprising because the vectors $ A_{\mu MN}$ gauge an 
abelian subalgebra of the gauge algebra.
On the other hand the $G_{\mu}^{M}$ vectors get their masses through 
the Scherk--Schwarz mechanism by eating some of the $g_{MN}$ scalars.

However we also note that the covariant derivative of the ${\cal A}_{MNP}$ 
scalars contains terms of the form
\begin{equation}
{\cal D}_{\mu} {\cal A}_{MNP} = \partial_{\mu} {\cal A}_{MNP}  + 
g_{MNPQ}G^{Q}_{\mu} + 3 \tau_{[MN}^{Q} A_{\mu P]Q} - 3 G_{\mu}^{Q} 
\tau_{Q[M}^{S} {\cal A}_{NP]S}\,,
\label{covDA}
\end{equation}
so that it also contributes to the vector boson masses.
Note however that this coupling is invariant under the shift symmetry 
$\delta A_{\mu MN} = \tau_{MN}^{Q} \xi_{\mu Q}$ because of the Jacobi 
identity $\tau_{[MN}^{S} \tau_{P]S}^{Q} = 0$.
Therefore the scalars ${\cal A}_{MNP}$ give mass to those vectors 
which are not eaten by the antisymmetric tensors.
Because of the flux term, the $G_{\mu}^{Q}$ vectors receive a 
contribution to the mass both from the metric scalars $G_{MN}$ and 
from the form scalars ${\cal A}_{MNP}$.
As expected, the physical mass will depend on both $\tau^{K}_{MN}$ 
and $g_{MNPQ}$.
This is the linearized analysis, the full non--linear gauge 
invariance is more complicated because of the non--abelian nature of 
the Scherk--Schwarz gauge group.

Let us now discuss the embedding of the gauge algebra in the 
4--dimensional ${\cal N} =8$ duality algebra.
The gauge field strengths and their duals lie in the $56$ of 
$E_{7(7)}$ and the full action of the algebra on them is described by 
a lower--triangular matrix
\begin{equation}
S = \left( 
\begin{array}{cc}
a&0\\
c& -a^{T}
\end{array}
\right),
\end{equation}
where $a$ is the part acting on the gauge fields described above and
$c$ acts on the nilpotent part of the splitting of the
$E_{7(7)}/SU(8)$ coset.
This latter can be constructed \cite{Solva} as
$Solv\left(E_{7(7)}/SU(8)\right) \supset Solv\left(GL(7)/SO(7)\right)
+ 7^{\prime-4} + 35^{-2}$ where the generators $X_M^{-4}$, $X^{MNP
-2}$ (dual to the scalars $A^{M+4}$, $A^{+2}_{MNP}$) setup the algebra
\begin{eqnarray}
[X_M,X_N] & = & [X_M,X^{MNP}] = 0, \\[2mm]
[X^{MNP},X^{QRS}] & = & \epsilon^{MNPQRST} X_T. 
\end{eqnarray}
Given the splitting $56\to 7^{-3}+21^{\prime-1}+7^{\prime+3}+21^{+1}$,
according to the field--strengths of the vectors $G_{\mu}^{M}$,
$A_{\mu MN}$ and of their duals, the explicit form of these matrices
is given by
\begin{equation}
a = \left(
\begin{array}{cc}
L^M{}_N& 0\\
\xi_{MNP}& 2L_{[M}^{P}\delta_{N]}^Q
\end{array}
\right), \qquad
c = \left(
\begin{array}{cc}
0&2\eta^{[P}\delta^{Q]}_{M}\\
2\eta^{[M}\delta^{N]}_P&\epsilon^{MNPQRST} \xi_{RST}
\end{array}
\right), 
\label{ac}
\end{equation}
where the $a$ matrix elements follow from (\ref{delG}) and 
(\ref{aprime})
\begin{equation}
\begin{array}{rcl}
L^M{}_{N}&=& \omega^{P} \tau^{M}_{NP},\\[2mm]
\xi_{MNP}&=& 3\tau_{[MN}^{S} \Sigma_{P]S} + g_{MNPQ} \omega^Q,
\end{array}
\label{LXI}
\end{equation}
and the $c$ matrix elements can be read from the linear axion
couplings in the Lagrangian \cite{Andrianopoli:2004sv} of the form
${\cal A}^{P} {\cal F}^{L}(G) \wedge {\cal F}_{PL}(A)$ and
$\epsilon^{MNPQRST} {\cal A}_{RST} {\cal F}_{MN}(A) \wedge {\cal
F}_{PQ}(A)$, where ${\cal A}^{P}$ are the scalars dual to ${\cal
A}_{\mu\nu M}$.
In fact this implies that $\eta^{M} = \epsilon^{MNPQRST} g_{NPQR}
\Sigma_{ST}$.

It is clear that both the Scherk--Schwarz group and the flux
generators act as a triangular subgroup of the electric group
$e_{7(7)}$ according to the algebra decomposition $e_{7(7)} \to sl(7)
+ so(1,1) + t_{7^{\prime-4}} + t_{7^{+4}} + t_{35^{-2}} +
t_{35^{\prime+2}}$, where the weights of the nilpotent generators are
given according to their $SL(7)^{so(1,1)}$ representation.
From this, we can also be more specific on the group which can be
effectively gauged in the 4--dimensional theory, which is any
7--dimensional subgroup of $SL(7)$ whose adjoint representation
coincides with the fundamental of $SL(7)$ plus any subset of the
$t_{7^{\prime-4}} + t_{35^{-2}}$ nilpotent generators which lie in the
electric subalgebra.

We note that the 28 vector potentials with positive $SO(1,1)$ weight
do not complete the 28--dimensional $SL(8)$ representation.
In order to do that we need an electromagnetic duality rotation
between the {\bf 7} and {\bf 7}$^\prime$ (or {\bf 21} and {\bf
21}$^{\prime}$) \cite{Cremmer:1979up}.
Therefore our gauge algebra cannot be obtained within the usual
$SL(8)$ covariant formulation of ${\cal N} = 8$ supergravity.
It is also different from the ``flat'' group of \cite{Cremmer:1979uq}
since this latter also has vector potentials which differ from ours
by electromagnetic duality rotations.  

\section{${\cal N} =4$ type II orientifolds with fluxes}
\label{orientsec}

We now pass to the discussion of the effective theories coming from
${\cal N} =4$ orientifold constructions of type II theories.
In the following we are first going to write down the general gauge
and Kaluza--Klein transformations of the various fields of type IIA/B
theory.
Then we will specify them for the various cases which correspond to
the ${\cal N} = 4$ orbifolds of
\cite{Angelantonj:2003rq,Angelantonj:2003up}.
Finally we will compute the commutators and interpret the results in
terms of the generators of the symmetries of the scalar manifold.

\subsection{General transformations}
\label{generaltransform}

As already stressed in the previous section, the vector fields coming 
from the metric maintain the same transformation rule as in 
(\ref{delG})
\begin{equation}
\delta G_{\mu}^M = \partial_\mu \omega^M - \tau_{NP}^M \omega^N G^P_\mu.
\end{equation}
the other fields instead depend on whether we are in type IIA or type 
IIB theory.

Let us start from type IIB supergravity.
In addition to the vectors coming from the metric, one also has up to
12 vectors from the Neveu--Schwarz and Ramond--Ramond 2--forms ${\cal
B}$ and ${\cal C}$ and up to 20 vectors coming from the 4--form $\hat
{\cal C}$.

The derivation of the transformation rules for the vectors coming from
the 2--forms follows straightforwardly from the application of the
procedure used in the previous section.
All the fields coming from the reduction of the 2--form fields are
subject to transformations under the relic of their 10--dimensional
gauge symmetry ${\cal B} \to {\cal B} + d\Lambda$, ${\cal C} \to {\cal
C} + d\Gamma$ as well as of the diffeomorphisms ${\cal B} \to {\cal B}
+ {\cal L}_{\omega} {\cal B}$, ${\cal C} \to {\cal C} + {\cal
L}_{\omega} {\cal C}$.
In order to take in the proper account also the fluxes we have to 
define
\begin{equation}
{\cal B} = b + \beta, \quad \hbox{and} \quad {\cal C} = c+ \sigma,
\label{decc}
\end{equation}
where $b$ and $c$ contain the expansions of the fluctuations over the
backgrounds and $d\beta = h = \frac{1}{3!} h_{MNP} e^M \wedge e^N
\wedge e^P$ and $d\sigma = f=\frac{1}{3!} f_{MNP} e^M \wedge e^N \wedge
e^P$ are the non--trivial fluxes.
The transformations of the 4--dimensional fluctuations is then
\begin{eqnarray}
\delta b &=& i_{\omega} db + d i_{\omega} b + d \Lambda + i_{\omega} 
h, 
\label{varB} \\[2mm]
\delta c &=& i_{\omega} dc + d i_{\omega} c + d \Gamma + i_{\omega} 
f.
\label{varC}
\end{eqnarray}
Moreover, in order to get 4--dimensional fields which transform 
covariantly with respect to the gauge transformations, one also 
employs the composite definitions \cite{Scherk:1979zr}
\begin{eqnarray}
B_{M\mu} & \equiv & {\cal B}_{M\mu} - {\cal B}_{MN}G^{N}_{\mu}\,, 
\label{BMmudef}\\[2mm]
B_{\mu\nu} & \equiv & {\cal B}_{\mu\nu}- {\cal B}_{M\nu}G^{M}_{\mu}
- {\cal B}_{\mu N}G^{N}_{\nu}+ {\cal B}_{MN} G^{M}_{\mu} G^{N}_{\nu}  
\,.
\label{Bmunu}\\[2mm]
C_{M\mu} & \equiv & {\cal C}_{M\mu} - {\cal C}_{MN}G^{N}_{\mu}\,, 
\label{CMmudef}\\[2mm]
C_{\mu\nu} & \equiv & {\cal C}_{\mu\nu}- {\cal C}_{M\nu}G^{M}_{\mu}
- {\cal C}_{\mu N}G^{N}_{\nu}+ {\cal C}_{MN} G^{M}_{\mu} G^{N}_{\nu}  
\,.
\label{Cmunu}
\end{eqnarray}
The resulting transformations of the vector fields are then
\begin{equation}
\begin{array}{rcl}
\delta B_{M\mu} &=& -\omega^{P}h_{PMN} G^{N}_{\mu}+\omega^{P} 
\tau^{S}_{PM} B_{S\mu} - \partial_{\mu}\Lambda_{M} -
\tau^S_{MN} \Lambda_{S} G^{N}_{\mu}\,,\\[2mm]
\delta C_{M\mu} &=& -\omega^{P}f_{PMN} G^{N}_{\mu}+\omega^{P} 
\tau^{S}_{PM} C_{S\mu} - \partial_{\mu}\Gamma_{M} -
\tau^S_{MN} \Gamma_{S} G^{N}_{\mu}\,.
\end{array}
\label{bprime}
\end{equation}

More care is required in order to obtain the correct transformation
rules of the vector fields coming from the 4--form potential $\hat
{\cal C}$.
The definition of its field--strength contains a non--trivial
Chern--Simons term, which reads 
\begin{equation}
G_5 = d \hat {\cal C} + {\cal B} \wedge d{\cal C} - {\cal C} \wedge
d{\cal B}.
\end{equation}
There are mainly two points of concern regarding this definition.
The first is that the field strength $G_5$ is not invariant under the
gauge transformations of ${\cal B}$ and ${\cal C}$ unless $\hat {\cal
C}$ transforms as $\hat {\cal C} \to \hat {\cal C} - \Lambda \wedge d
{\cal C} + \Gamma \wedge d {\cal B}$.
The second comes from the proper definition of the fluctuations above
the expectation value, which we now want to be non--trivial because of
the addition of fluxes.
Actually, since we want the field--strength $G_5$ to be $y^M$
independent, the $y^M$ dependent contributions coming from ${\cal B}$
and ${\cal C}$ must cancel.
This problem is solved if we define $\hat {\cal C} \equiv \hat c +
\gamma - \beta \wedge c + \sigma \wedge b$, where we used (\ref{decc})
and $\hat c$ denotes the real 4--form fluctuations.
By doing so, and requiring that the 5--form flux $g$ follows from $g =
d\gamma + \beta \wedge f - \sigma \wedge h$, one gets an expression
for $G_5$ which is indeed $y^M$ independent:
\begin{equation}
G_5 = d \hat c + b \wedge dc - c \wedge db + 2( b \wedge f - c \wedge
h) + g.
\label{G5}
\end{equation}
The transformation rules of the 4--form fluctuations can then be
derived by recalling that $\delta \hat {\cal C} = \delta \hat c -
\beta \wedge \delta c + \sigma \wedge \delta b = i_\omega d \hat {\cal
C} + d i_\omega \hat {\cal C} - \Lambda \wedge d {\cal C} + \Gamma
\wedge d {\cal B} + d \Sigma$.
After some algebra one gets that
\begin{equation}
\delta \hat c = i_{\omega}d\hat c + d i_{\omega} \hat c + i_{\omega}g 
+ b \wedge  i_{\omega} f - c \wedge  i_{\omega}h -\Lambda \wedge (dc 
+ 2f) + \Gamma \wedge (db + 2 h) + d\Sigma.
\label{deltahatC}
\end{equation}

As happened for the 2--forms, the proper covariant fields are not 
directly the reduced 4--form fields, but rather certain their 
combinations.
We are not going to discuss here the complete structure of all the 
redefinitions and transformations, but focus only on the vector 
fields, which are those needed in order to obtain the gauge algebra.
From (\ref{deltahatC}) one obtains that the vector fields transforming 
covariantly are
\begin{equation}
C_{\mu MNP}=  \hat c_{\mu MNP} +\hat c_{MNPQ} \, 
G^{Q}_{\mu}+ 3 B_{\mu[M}{\cal C}_{NP]}-3C_{\mu[M} {\cal B}_{NP]}\,,
\label{C4def}
\end{equation}
and their variation is 
\begin{equation}
\begin{array}{rcl}
\delta C_{\mu MNP} & = & \partial_{\mu} \Sigma_{MNP} +3 \,\omega^{R} 
\, \tau^{S}_{R[M}\, C_{|\mu S| NP]} - 3 \tau^{S}_{[MN}\Sigma_{|\mu S|P]}
 - 3 G_{\mu}^{Q} \tau_{Q[M}^{S} 
\Sigma_{|S|NP]}  \\[2mm]
&-& \Lambda_\mu f_{MNP}  - 6 \Lambda_{[M}f_{NP]Q}G_{\mu}^Q 
+ \Gamma_\mu h_{MNP}  + 6 \Gamma_{[M}h_{NP]Q}G_{\mu}^Q \\[2mm]
&+& \omega^{R} g_{RMNPQ} G^{Q}_{\mu}
+  6 \omega^{R} B_{\mu [M}f_{NP]R}
-6  \omega^{R} C_{\mu [M}h_{NP]R}.
\end{array}
\label{deltaCmuMNP}
\end{equation}

We see also in this transformation some unusual terms related to the
gauge transformations of the tensor fields present in the reduced 
theory.
As for the M--theory case analyzed previously, this means that the
generic gauge algebra of the effective theory is actually a free
differential algebra, involving also these tensor gauge
transformations.
The orientifold projections in some cases help us in getting rid of
the extra fields so that one obtains ordinary Lie algebras in the
4--dimensional theory.
When this does not happen, the closure of the ordinary Jacobi
identities shows the conditions needed for a real interpretation in 
terms of ordinary gauge algebras and therefore we will give them 
explicitly in the following.

\subsubsection{Type IIA transformations}

In type IIA supergravity the discussion follows the same path.
The difference is only in the field content.
There is a one--form ${\cal A} = a + \sigma$ whose 4--dimensional
gauge field fluctuation $A_\mu \equiv a_\mu + a_M G^M_\mu$ transforms
as
\begin{equation}
\delta A_{\mu} = \partial_{\mu} \lambda - \omega^{N} f_{NM} 
G^{M}_{\mu}.
\label{deltaA}
\end{equation}
Next, there is the Neveu--Schwarz 2--form ${\cal B}$ which transforms
as (\ref{bprime}).
Finally there is a 3--form potential ${\cal C}$, whose transformation
law depends again on its field--strength definition $G_4 = d{\cal C} +
{\cal B} \wedge d{\cal A}$.
As for the type IIB 5--form $G_5$, we get a $y^M$--independent
definition for $G_4$ by setting ${\cal C} = c + \gamma + \sigma \wedge
b$ and $d\gamma + \sigma h = g$:
\begin{equation}
G_4 = d c + a \wedge db + a \wedge h + b \wedge f+ g.
\label{G4}
\end{equation}
The resulting transformation law is
\begin{equation}
\delta c = i_{\omega} dc + di_{\omega}c + i_{\omega} g  + i_{\omega}f 
\wedge b-\lambda ( db + h) + \Lambda \wedge f + d \Sigma.
\label{delC3}
\end{equation}

Once again, one has to obtain the proper definition of the
4--dimensional degrees of freedom, which, for what concerns the vector
fields, are $C_{\mu MN} = {\cal C}_{\mu MN} - {\cal C}_{QMN} G_\mu^Q -
A_\mu {\cal B}_{MN}$.
These fields transform as
\begin{equation}
\begin{array}{rcl}
\delta C_{\mu MN} &=& \partial_{\mu}\Sigma_{MN} +  2 \omega^Q 
\,\tau^S_{Q[M} C_{|S|N]\mu} + \Lambda_{\mu} f_{MN}  \\[2mm]
&-& \omega^Q g_{QMNP} 
G^{P}_{\mu} - \tau_{MN}^S \Sigma_{\mu S}- \tau_{MN}^S  \Lambda_S A_{\mu}
 - 2 G_{\mu}^P 
\tau^S_{P[M}\Sigma_{|S|N]} \\[2mm]
&-& 2 B_{\mu[M}f_{N]P}\omega^P- 2 \Lambda_{[M}f_{N]P}G_\mu^P + 
\lambda h_{MNP}G_\mu^P - A_\mu \omega^P h_{PMN}.
\end{array}
\label{deltaCmuMN}
\end{equation}
From this transformation we deduce that also in type IIA reductions
the gauge fields will generically transform under tensor field gauge
transformations.
Once again, in the following we will give the conditions needed for an
interpretation of the algebra of these transformations in terms of
ordinary gauge algebras.

\subsection{IIB Scherk--Schwarz compactification with fluxes and
O9--planes}

Now that we have the general form of the transformation laws for the
various vector fields with respect to both Kaluza--Klein and gauge
transformations, we can specialize the results of the previous section
for the orientifolded IIB theories which yield ${\cal N} = 4$ effective
supergravity.
To obtain these theories one has to consider the various orientifold 
projections on the fields and space--time, which imply a reduced 
spectrum with respect to that considered in the previous section.
Neglecting the D--branes degrees of freedom, we will always have a 
total of 12 vector fields, which however originate from 
different 10--dimensional fields, depending on the orbifold action.
We will also consider contributions coming from both physical fluxes 
as well as from those which survive the orientifold projections 
despite their potentials do not.

We start again with a simple case, which however presents most of the
relevant features which appear in the general one.
This is the case of Scherk--Schwarz compactifications of type IIB
theory when also D9--branes and O9--planes are present.
For such compactification the orbifold projection does not touch the
internal coordinates, which from now on we will label according to the
orbifold action: first those which are invariant and than those which
are acted on by the orbifold projection $M = \{i,a\}$.
For O9--planes we will have $M = i = 1,\ldots,6$.
The same projection implies that the vector field content is reduced
to 6 vectors coming from the metric and 6 coming from the
Ramond--Ramond 2--form:
\begin{equation}
G_{\mu}^{i}, \ C_{i\mu}.
\label{vecD9O9}
\end{equation}
In addition, one can turn on a single type of flux, the Ramond--Ramond
3--form\footnote{The effect of this flux to the effective theory was
not considered in \cite{Angelantonj:2003rq,Angelantonj:2003up}, where
only ``unphysical'' fluxes were used.}
\begin{equation}
f_{ijk}.
\label{fluxD9O9}
\end{equation}
As a last ingredient, since we are going to perform a Scherk--Schwarz
reduction, we also have the possibility to introduce geometrical 
fluxes, which correspond to the Scherk--Schwarz structure constants
\begin{equation}
\tau^{i}_{jk}.
\label{torsionD9O9}
\end{equation}
Given this setup, the transformations of the vector fields are 
\begin{equation}
\delta G_{\mu}^{i } = \partial_{\mu}\omega^{i} - \tau^{i}_{jk} 
\omega^{j} G^{k}_{\mu}\,,
\label{varG9}
\end{equation}
for the vectors coming from the metric and
\begin{equation}
\delta C_{i\mu} =  -\omega^k f_{kij}G_\mu^j + \omega^{k} 
\tau^{j}_{ki} C_{j\mu} - \partial_{\mu}\mu_{i} - 
\tau^{j}_{ik}\mu_{j} G_{\mu}^k,
\label{varC9}
\end{equation}
for the variation of the Ramond--Ramond 2--form.

We can now deduce the structure of the gauge algebra of the effective 
theory assigning to the 12 gauge generators the following labels
\begin{equation}
T_{A} = \{ Z_i,  Y^i\},
\label{gen5}
\end{equation}
matching the parameters $\{ \omega^i,  \mu_i\}$ respectively.
Considering commutators of (\ref{varG9}) and (\ref{varC9}) we can 
again obtain the commutators of the resulting gauge algebra which 
reads
\begin{equation}
\begin{array}{rcl}
[Z_i, Y^j] &=& \tau^j_{ik} \, Y^k,\\[2mm]
[Z_i, Z_j] &=& f_{ijk}Y^k + \tau^k_{ij} Z_k.
\end{array}
\label{algebra9}
\end{equation}
In deriving this algebra we also obtain some constraints on the 
structure constants:
\begin{equation}
\tau_{ij}^j = 0, \quad \hbox{ and } \tau_{[ij}^m f_{kl]m}=0.
\label{consttau}
\end{equation}
These allow the closure of the Jacobi identities and have again the 
interpretation of the invariance of the volume of the internal space 
and closure of the Bianchi identity of the Ramond--Ramond 3--form 
flux $f$.

It is possible at this stage to account for the embedding of this 
gauge algebra in the duality group.
In this case the duality matrix (\ref{Sduality}) contains only the 
matrix $a$, whereas $c = 0$.
This last term is absent because in type IIB theory with 9--branes 
and O9--planes the only axion coupling is 
$SL(6)$ invariant, $c {\cal F}^{M}(G) \wedge \tilde {\cal 
F}_{M}(\tilde C)$, so that we only have a global symmetry with 
respect to the external $SU(1,1)$ factor of the duality group.

The other term is fixed by the $SO(6,6)$ action on the vector fields
of the theory, which we group as $\{G_\mu^i,C_{\mu i}\}$.
This action is then represented by the SO(6,6) matrix
\begin{equation}
a = \left( 
\begin{array}{cc}
L & 0 \\ X & -L^T
\end{array}
\right),
\label{so66matr}
\end{equation}
where $L = \tau^{i}_{jk} \omega^k$ and $X = \omega^k f_{kij} +
\tau_{ij}^k \mu_k$.
It is then clear from this explicit representation that the
Scherk--Schwarz group acts as a triangular subgroup of $SO(6,6)$
according to the decomposition $so(6,6) \to o(1,1) + sl(6) +
t_{15^\prime}^{-2} + t_{15}^{+2}$, which corresponds to a solvable
decomposition of $so(6,6)$.
In detail, the matrix $L$ is an $sl(6)$ matrix and the $X$
coefficients are in the representation {\bf 15}$^{\prime-2}$ of
$SL(6)^{o(1,1)}$.
As we see, this structure is not changed by the addition of fluxes.
This same embedding gives us also some more information on the
possible gauge group of the effective theory represented by the
algebra (\ref{algebra9}).
The representation (\ref{so66matr}) implies that we can gauge any
6--dimensional subgroup of $SL(6)$ and all the Peccei--Quinn
symmetries of the scalar manifold corresponding to the nilpotent
generators in the $t_{15}^{+2}$, which are those associated to the
Ramond--Ramond scalars $C_{ij}$.
This pattern will be followed also by the other examples where the
nilpotent part is always given by a $15+(p-3)(9-p)$ dimensional
subalgebra of $Solv(so(6,6))$ related to the Peccei--Quinn symmetries
of the scalar manifold.

\subsection{IIB Scherk--Schwarz compactification with fluxes and
O7--planes}

We will now detail another interesting case, which presents some more
features than the previous: the type IIB Scherk--Schwarz
compactification with fluxes and O7--planes.
Then we will list the results for all the other cases.

When O7--planes are present, the orientifold projection splits the
internal space in 4 invariant coordinates $i$ and 2 which are acted
upon $a$.
The vector field content is given by 4 vectors coming from the metric,
2 from the Neveu--Schwarz 2--form, 2 from the Ramond--Ramond 2--form
and 4 from the 4--form:
\begin{equation}
G_{\mu}^{i}, \ B_{a\mu}, \ C_{a\mu} \ C_{\mu ijk}.
\label{vecD7O7}
\end{equation}
The allowed fluxes are
\begin{equation}
h_{ija}, \ f_{ija}, \ g_{ijkab},
\label{fluxD7O7}
\end{equation}
and in this case also some of the geometrical fluxes are projected 
away from the orbifold projection.
The surviving ones are
\begin{equation}
\tau^{i}_{ab}, \ \tau^{i}_{jk}, \ \tau^{a}_{bi}.
\label{torsionD7O7}
\end{equation}

If we now specialize the transformations of the vector fields listed 
in section \ref{generaltransform}, we get
\begin{equation}
\delta G_{\mu}^{i } = \partial_{\mu}\omega^{i} - \tau^{i}_{jk} 
\omega^{j} G^{k}_{\mu}\,,
\label{varG7}
\end{equation}
for the vectors coming from the metric,
\begin{equation}
\delta B_{a\mu} = - \omega^{i} h_{iaj} G_{\mu}^{j} + \omega^{i} 
\tau^{b}_{ia} B_{b\mu} - \partial_{\mu}\Lambda_{a} - 
\tau^{b}_{ai}\Lambda_{b} G_{\mu}^i,
\label{varB7}
\end{equation}
for the variation of the Neveu--Schwarz 2--form,
\begin{equation}
\delta C_{a\mu} = - \omega^{i} f_{iaj} G_{\mu}^{j} + \omega^{i} 
\tau^{b}_{ia} C_{b\mu} - \partial_{\mu}\mu_{a} - 
\tau^{b}_{ai}\mu_{b} G_{\mu}^i,
\label{varC7}
\end{equation}
for the variation of the Ramond--Ramond 2--form and
\begin{equation}
\delta C_{\mu ijk} = \partial_{\mu} \Sigma_{ijk} + 3 \omega^{l} 
\tau^m_{l[i} C_{|\mu m|jk]} - 3 G_{\mu}^l \tau 
^m_{l[i}\Sigma_{|m|jk]} + 3\tau_{[ij}^{l} \Sigma_{\mu k]l}
\label{varhatC7}
\end{equation}
for the 4--form.

Once more we can extract the structure of the gauge algebra from the 
above transformations by considering their commutators.
Assigning to the gauge generators the labels
\begin{equation}
T_{A} = \{ Z_i, X^a, Y^a, W_i \},
\label{gen7}
\end{equation}
matching the parameters $\{ \omega^i, \Lambda_a, \mu_a, 
\epsilon^{ijkl} \Sigma_{jkl} \}$, we find the following structure for 
the gauge algebra 
\begin{equation}
\begin{array}{rcl}
[Z_i, X^a] &=& \tau^a_{bi} \, X^b,\\[2mm]
[Z_i, Y^a] &=& \tau^a_{bi} \, Y^b,\\[2mm]
[Z_i, W_j] &=& \tau^k_{ij} \, W_k - \tau^k_{ik} W_j,\\[2mm]
[Z_i, Z_j] &=& h_{ija} X^a + f_{ija} Y^a + \tau^k_{ij} Z_k.
\end{array}
\label{algebra7}
\end{equation}
To get these commutators we need to impose some constraints on the 
geometrical fluxes
\begin{equation}
\tau^S_{[MN} \tau^Q_{P]S} = 0.
\label{torcon1}
\end{equation}
At the same time there are constraints affecting also the real fluxes
\begin{equation}
\tau^b_{a[i} h_{jk]b} = 0, \quad \tau^b_{a[i} f_{jk]b} = 0,
\label{con1}
\end{equation}
which are once more a consequence of the 10--dimensional Bianchi 
identities.
It is interesting to notice that the Jacobi identities in this case 
impose a further constraint
\begin{equation}
\tau_{im}^m = 0.
\end{equation}
Though this may seem strange, it just states that not only the total
volume is invariant under diffeomorphisms (which yields the known
$\tau_{MN}^N = 0$), but also the volume of the D7 brane should remain
fixed under the surviving diffeomorphisms.
This same constraint has also an interpretation as the tracelessness
condition for $sl(4)$ matrices when discussing the embedding of this
algebra in the duality one.
Together with $\tau_{ia}^{a} = 0 = \tau_{ab}^{i}$, this condition is 
needed in order to interpret the above algebra as an ordinary Lie 
algebra, which is a necessary condition to discuss its embedding in 
the standard full duality algebra.
When these conditions are not satisfied, one has to employ the full 
free differential algebra containing also the tensors $C_{\mu\nu ij}$.

For this case we will not discuss the full form of the duality action
(\ref{Sduality}), but only show the form of the $a$ matrix, which is 
the one containing the information of the gauged group.
The Scherk--Schwarz and fluxes action on the vectors can be described
according to the decomposition 
\begin{equation}
\begin{array}{rcl}
so(6,6) &\to& o(1,1)^2 + sl(4) + sl(2) + t_{(4,2)^{(1,0)}} + 
t_{(4,2)^{(1,1)}} + t_{(6,1)^{(2,1)}}
+ t_{(1,1)^{(0,1)}}\\[2mm]
&&+t_{(4,2)^{(-1,0)}} + 
t_{(4,2)^{(-1,-1)}} + t_{(6,1)^{(-2,-1)}}
+ t_{(1,1)^{(0,-1)}},
\end{array}
\label{decoo66}
\end{equation}
where the nilpotent generators $t$ are labelled according to their
$[SL(4)\times SL(2)]^{o(1,1)^2}$ representations.
The Scherk--Schwarz action in particular contains elements gauging the
$sl(4) + sl(2) + t_{(4,2)^{(1,0)}} + t_{(4,2)^{(1,1)}} +
t_{(6,1)^{(2,1)}}$ part of the algebra and the fluxes contribute to the
$t_{(4,2)^{(1,0)}} + t_{(4,2)^{(1,1)}}$ part:
\begin{equation}
a= \left(
\begin{array}{cccc}
-\omega^k \tau_{kj}^i & 0 & 0& 0 \\
\tau_{aj}^b \Lambda_b + \omega^k h_{kaj}& \omega^k \tau_{kb}^a& 0& 0\\
-\tau_{jk}^{m} \Sigma^k& 0 & \omega^k \tau_{ki}^j& 0\\
\tau_{aj}^b \mu_b + \omega^k f_{kaj}&0 &0 & \omega^k \tau_{ka}^b
\end{array}
\right).
\end{equation}
We therefore see that all the Peccei--Quinn symmetries, but the one of
the scalar resulting from the 10--dimensional axion $C_0$ and of $\hat 
{\cal C}_{ijkl}$, can be gauged.

\subsection{The other cases}
\label{othercasessec}

Now that we have discussed in detail some examples explaining also 
the derivation of the algebras, we will simply list the results for 
the other possible ${\cal N} = 4$ orientifolds in both type IIB and 
IIA.

\subsubsection{Type IIB with D5--branes and O5--planes}

The vector field content is given by
\begin{equation}
G_{\mu}^{i}, \ B_{a\mu}, \ C_{i\mu} \ C_{\mu abc},
\label{vecD5O5}
\end{equation}
and the allowed fluxes are
\begin{eqnarray}
&h_{abc}, \ h_{ija}, \ f_{iab}, \ g_{ijabc},&
\label{fluxD5O5}\\[2mm]
&\tau^{i}_{ab}, \ \tau^{i}_{jk}, \ \tau^{a}_{bi}.&
\label{torsionD5O5}
\end{eqnarray}
The transformations of the vector fields read
\begin{eqnarray}
\delta G_{\mu}^{i } &=& \partial_{\mu}\omega^{i} - \tau^{i}_{jk} 
\omega^{j} G^{k}_{\mu}\,,
\label{varG5}\\[2mm]
\delta B_{a\mu} &=& - \omega^{i} h_{iaj} G_{\mu}^{j} + \omega^{i} 
\tau^{b}_{ia} B_{b\mu} - \partial_{\mu}\Lambda_{a} - 
\tau^{b}_{ai}\Lambda_{b} G_{\mu}^i,
\label{varB5}\\[2mm]
\delta C_{i\mu} &=&  \omega^{k} 
\tau^{j}_{ki} C_{j\mu} - \partial_{\mu}\mu_{i} - 
\tau^{j}_{ik}\mu_{j} G_{\mu}^k,
\label{varC5}\\[2mm]
\delta C_{\mu abc} &=& \partial_{\mu} \Sigma_{abc} - 3 G_{\mu}^i \tau 
^d_{i[a}\Sigma_{|d|bc]} - 6\Lambda_{[a}f_{bc]i} G_\mu^i + 2\Gamma_\mu 
h_{abc} \\[2mm]
&+&3 \omega^i 
\tau^d_{i[a} C_{|\mu d|bc]} 
+ \omega^i g_{iabcj} G_{\mu}^j +  6 \omega^i B_{\mu[a}f_{bc]i} + 
3\tau_{[ab}^{i} \Sigma_{\mu c]i}.
\label{varhatC5}
\end{eqnarray}
The resulting gauge algebra is 
\begin{equation}
\begin{array}{rcl}
[Z_i, X^a] &=&  \tau^a_{bi} \, X^b+6 f_{bci}\epsilon^{abcd}W_d,\\[2mm]
[Z_i, Y^j] &=& \tau^j_{ki} \, Y^k,\\[2mm]
[Z_i, W_a] &=&  \tau^b_{ia} \, W_b,\\[2mm]
[Z_i, Z_j] &=&  h_{ija} X^a -g_{ijabc} \epsilon^{abcd} W_d + 
\tau^k_{ij} Z_k,
\end{array}
\label{algebra5}
\end{equation}
where we assigned the following labels to the gauge generators 
\begin{equation}
T_{A} = \{ Z_i, X^a, Y^i, W_a \},
\label{gen5b}
\end{equation}
matching the parameters $\{ \omega^i, \Lambda_a, \mu_i, 
\epsilon^{abcd} \Sigma_{bcd} \}$ respectively.

It is interesting to note that in this case there is no non--vanishing
commutator between $X^a$ and $Y^i$ despite the fact that this could
close on the $W_a$ generators using $\tau_{ab}^i \epsilon^{abcd}$ as
structure constants.
The straightforward computation of the above commutators gives also an
extra term in the commutator between $Z$ and $W$ of the form $\tau_{i
b}^b W_a$, but, as above, this has to vanish for an interpretation of
these commutators as those of an ordinary Lie algebra.
This follows from the closure of the Jacobi identities and can be
interpreted as the requirement that the volumes of the D5 branes be
invariant under diffeomorphisms.
Obviously, when the Scherk--Schwarz terms are turned off, these
results match those in \cite{Angelantonj:2003rq,Angelantonj:2003up},
but for some new contribution coming from the 5--form flux which was
not noticed before.

The closure of the algebra requires that the 3--form flux satisfies
\begin{equation}
2\tau_{i[a}^d f_{b]j d} +\tau_{ij}^k f_{abk} = 0.
\label{ttf}
\end{equation}
This has the interpretation as the Bianchi identity of the 3--form 
flux in 10 dimensions.
A similar expression for the 5--form flux is not obtained because of 
the index structure, i.e. $[ijk] = 0$.
The same argument also tells us that the 3--form fluxes are such that
\begin{equation}
f \wedge h =  0,
\label{hwedgef}
\end{equation}
which is compatible with the tadpole cancellation conditions for D5 
branes.

In this case the gauge algebra acts according to the decomposition of 
the duality group as
\begin{equation}
\begin{array}{rcl}
so(6,6) &\to& o(1,1)^2 + sl(2) + sl(4) + t_{(1,6)^{(0,1)}} + 
t_{(2,4)^{(1,0)}} + t_{(2,\bar 4)^{(1,1)}} + 
t_{(1,1)^{(2,1)}}\\[2mm]
&&  t_{(1,\bar 6)^{(0,-1)}} + 
t_{(2,4)^{(-1,0)}} + t_{(2,\bar 4)^{(-1,-1)}} + 
t_{(1,1)^{(-2,-1)}},
\end{array}
\label{dec66bis}
\end{equation}
where again the nilpotent generators $t$ are labelled using their
representations under $[SL(2)\times SL(4)]^{O(1,1)^2}$ and the
Scherk--Schwarz action contains the $sl(2) + sl(4) + t_{(2,4)^{(1,0)}}
+ t_{(2,\bar 4)^{(1,1)}} + t_{(1,1)^{(2,1)}}$ generators.
In this case the fluxes do not simply overlap with the Scherk--Schwarz
action, but allow also for the gauging of the remaining Peccei--Quinn
symmetries, since the $\omega^i f_{ibc}$ terms are in the
$t_{(1,\bar 6)^{(0,-1)}}$ and couple to the generators in the
$t_{(1,6)^{(0,1)}}$.

\subsubsection{Type IIB with D3--branes and O3--planes}

The vector field content is given by
\begin{equation}
 B_{a\mu}, \ C_{a \mu},
\label{vecD3O3}
\end{equation}
and the allowed fluxes are
\begin{equation}
h_{abc}, \ f_{abc}.
\label{fluxD3O3}
\end{equation}
In this case no geometrical fluxes are allowed because of the
orientifold projection acting on the internal coordinates.
The resulting gauge algebra is therefore given by purely abelian
factors as noted in
\cite{Frey:2002hf,Kachru:2002sk,Angelantonj:2003rq}.
It should be stressed the role of the fluxes in this case.
Despite the fact that we obtain an effective theory which is abelian
and therefore $a=0$, the addition of fluxes still implies that there
is a gauging of some of the symmetries of the scalar manifold.
In fact also the action of the duality group is non--trivial on the
gauge field--strengths.
The $c$ terms in the matrix (\ref{Sduality}) are non--vanishing
\cite{DAuria:2003jk} and can be described by
\begin{equation}
c = \left(
\begin{array}{cc}
0 & h^{abc} \Lambda_c +  f^{abc} \Gamma_c  \\
-h^{abc} \Lambda_c -  f^{abc} \Gamma_c  & 0
\end{array}
\right).
\label{cd3}
\end{equation}

\subsubsection{Type IIA with D8--branes and O8--planes}

The vector fields surviving the orbifold projections are
\begin{equation}
G_{\mu}^{i}, \ A_{\mu}, \ C_{i9\mu}, \ B_{9\mu}.
\label{vecD8O8}
\end{equation}
The allowed fluxes are
\begin{eqnarray}
&f_{ij}, \ h_{ij9},  \ g_{ijk9},&
\label{fluxD8O8}\\[2mm]
&\tau^{i}_{jk}, \quad \tau^9_{9i},&
\label{torsionD8O8}
\end{eqnarray}
where the last term is forced to vanish by the invariance of the
volume of the D8 brane under the surviving diffeomorphisms.
The transformations of the vector fields are 
\begin{eqnarray}
\delta G_{\mu}^{i } &=& \partial_{\mu}\omega^{i} - \tau^{i}_{jk} 
\omega^{j} G^{k}_{\mu}\,,
\label{varG8}\\[2mm]
\delta A_{\mu} &=& \partial_\mu \lambda - f_{ij} \omega^i G_\mu^j,
\label{varC8}\\[2mm]
\delta B_{9 \mu} &=&  -\omega^i h_{i9j}G^j_\mu -\partial_{\mu} \Lambda_9,
\label{varB8}\\[2mm]
\delta C_{\mu i9} &=& \partial_\mu\Sigma_{i9}  -
G_\mu^j \tau_{ji}^k \Sigma_{k9}+ \omega^k \tau_{ki}^j 
C_{\mu j9} + \omega^k g_{i9jk} G_\mu^j \\[2mm]
&+& \lambda h_{i9j}G_{\mu}^j 
-A_{\mu} \omega^k h_{ki9} 
- B_{9\mu} f_{ij} \omega^j + \Lambda f_{ij} G_\mu^j - \tau_{i9}^{9} 
\Sigma_{\mu 9}.
\label{varC8bis} 
\end{eqnarray}
The resulting gauge algebra is then
\begin{equation}
\begin{array}{rcl}
[Z_i, X] &=& f_{ij} W^j, \\[2mm]
[Z_i, Y] &=& h_{ij9} W^j, \\[2mm]
[Z_i, W^j] &=& -\tau_{ik}^j W^k, \\[2mm]
[Z_i, Z_j] &=& f_{ij}Y + h_{ij9} X + g_{9ijk}W^k + \tau_{ij}^k Z_k,
\end{array}
\label{algebra8}
\end{equation}
where the gauge generators 
\begin{equation}
T_{A} = \{ Z_i, Y, X, W^i \},
\label{gen8}
\end{equation}
correspond to the parameters $\{ \omega^i, \lambda, \Lambda_9,
\Sigma_{i9} \}$.
Once more we point out that these algebra reduces to the one of
\cite{Angelantonj:2003up} when the geometrical fluxes are turned off,
but for the first commutator and the first term in the last commutator
which were not noticed there.
The Jacobi identities imply the following constraints on the structure
constants which can be again interpreted as the Bianchi identities in
10 dimensions:
\begin{equation}
\tau^l_{[ij}h_{k]9l} =0, \quad 
\tau^m_{[ij}g_{kl]9m} = f_{[ij}h_{kl]9}.
\label{fluxcond8}
\end{equation}
Moreover, we did not include terms of the type $\tau_{i9}^{9}$ and
$\tau_{ij}^{j}$ which vanish when considering an ordinary Lie algebra,
but may be present in the full free--differential algebra.
The action of the gauge group is embedded in the decomposition of the
duality group as
\begin{equation}
so(6,6) \to o(1,1)^2 + sl(5) + t_{5^{(0,1)}} + t_{5^{(1,0)}} + 
t_{10^{(1,1)}} + t_{\bar 5^{(0,-1)}} + t_{\bar 5^{(-1,0)}} + 
t_{\overline{10}^{(-1,-1)}}.
\end{equation}
The Scherk--Schwarz reduction gauges a subgroup of $sl(5) +
t_{10^{(1,1)}}$ while the fluxes gauge $t_{5^{(0,1)}} +
t_{5^{(1,0)}}+t_{10^{(1,1)}}$.

\subsubsection{IIA with D6--branes and O6--planes}

The vector fields are
\begin{equation}
G_{\mu}^{i}, \ B_{a\mu}, \ C_{ij\mu} \ C_{\mu ab},
\label{vecD6O6}
\end{equation}
and the allowed fluxes are
\begin{eqnarray}
&h_{abc}, \ h_{ija}, \ f_{ia}, \ g_{ijab},&
\label{fluxD6O6}\\[2mm]
&\tau^{i}_{ab}, \ \tau^{i}_{jk}, \ \tau^{a}_{bi}.&
\label{torsionD6O6}
\end{eqnarray}
The transformations of the vector fields read
\begin{eqnarray}
\delta G_{\mu}^i &=& \partial_{\mu}\omega^{i} - \tau^{i}_{jk} 
\omega^{j} G^{k}_{\mu}\,,
\label{varG6}\\[2mm]
\delta B_{a \mu} &=&  -\omega^i h_{iaj} G^j_{\mu} +\omega^i \tau^b_{ia} 
B_{b\mu}- \partial_{\mu}\Lambda_a - \tau_{ai}^b \Lambda_b G^i_{\mu},
\label{varB6}\\[2mm]
\delta C_{\mu ij} &=& \partial_{\mu}\Sigma_{ij} +  2 \omega^l 
\,\tau^k_{l[i} C_{|k|j]\mu}  - 2 G_{\mu}^l 
\tau^k_{l[i}\Sigma_{|k|j]} -\tau_{ij}^{k}\Sigma_{\mu k},\\[2mm]
\delta C_{\mu ab} &=& \partial_{\mu}\Sigma_{ab} +  2 \omega^i 
\,\tau^c_{i[a} C_{|c|b]\mu}  - \omega^i g_{iabj} 
G^{j}_{\mu} - 2 G_{\mu}^i 
\tau^c_{i[a}\Sigma_{|c|b]} \\[2mm]
&+&2\Lambda_{[a}f_{b]j}G^j_{\mu}- 2 \omega^i B_{\mu[a} 
f_{b]i} - \tau_{ab}^{i}\Sigma_{\mu i}.
\label{varC6}
\end{eqnarray}
We can assign to the gauge generators the following labels
\begin{equation}
T_{A} = \{ Z_i, X^a, W^{ij}, K^{ab} \},
\label{gen6}
\end{equation}
matching the parameters $\{ \omega^i, \Lambda_a, 
\Sigma_{ij}, \Sigma_{ab} \}$ to obtain the following gauge algebra 
\begin{equation}
\begin{array}{rcl}
[Z_i,X^a] &=& 2 f_{ic} K^{ca}+\tau_{bi}^a X^b,\\[2mm]
[Z_i,W^{jk}] &=& 2\tau_{il}^{[j}W^{k]l},\\[2mm]
[Z_i,K^{ab}] &=& 2\tau_{ic}^{[a} K^{b]c},\\[2mm]
[Z_i,Z_j] &=& h_{ija} X^a + g_{ijab} K^{ab} + \tau_{ij}^k Z_k.
\end{array}
\label{algebra6}
\end{equation}
Once more we have some constraints on the structure constants:
\begin{equation}
\tau^k_{ij}f_{ak}+2\tau^b_{a[i}f_{j]b} =0, \quad 
2\tau^c_{[a|[i}g_{jk]|b]c}-\tau^l_{[ij}g_{k]ab l} = 2 h_{ij[a}f_{b]k}.
\label{fluxcond6}
\end{equation}
The action of the gauge group is embedded in the decomposition of the 
duality group as
\begin{equation}
\begin{array}{rcl}
so(6,6) &\to& o(1,1)^2 + sl(3) +sl(3) + t_{(1,3)^{(0,1)}} + 
t_{(3,3)^{(1,0)}} + t_{(3, \bar 3)^{(1,1)}} + 
t_{(\bar 3, 1)^{(2,1)}}\\
&&+ t_{(1,\bar 3)^{(0,-1)}} + 
t_{(\bar 3,\bar 3)^{(-1,0)}} + t_{(\bar 3,  3)^{(-1,-1)}} + 
t_{(3, 1)^{(-2,-1)}}.
\end{array}
\end{equation}
The Scherk--Schwarz reduction gauges a subgroup of $sl(3) + sl(3) +
t_{(3,3)^{(1,0)}} + t_{(3, \bar 3)^{(1,1)}} + t_{(1,3)^{(0,1)}}$ while
the fluxes gauge a $ t_{(1,3)^{(0,1)}}+ t_{(3,3)^{(1,0)}} + t_{(\bar
3, 1)^{(2,1)}}$.

\subsubsection{IIA with D4--branes and O4--planes}

The surviving vector fields are
\begin{equation}
G_{\mu}^{4}, \ B_{a\mu}, \ C_{\mu},  \ C_{4a\mu},
\label{vecD4O4}
\end{equation}
and the allowed fluxes are
\begin{eqnarray}
&h_{abc}, \ f_{ab}, \ g_{4abc}.&
\label{fluxD4O4}\\[2mm]
&\tau^{4}_{ab},  \ \tau^{a}_{b4}.&
\label{torsionD4O4}
\end{eqnarray}
The transformations of the vector fields are 
\begin{eqnarray}
\delta G_{\mu} &=& \partial_{\mu}\omega\,,
\label{varG4}\\[2mm]
\delta A_{\mu} &=& \partial_\mu \lambda,
\label{varC4}\\[2mm]
\delta B_{a \mu} &=&  \omega \tau^b_{4a} B_{b\mu} - \partial_\mu 
\Lambda_a - \tau_{a4}^b \Lambda_{b} G_{\mu},
\label{varB4}\\[2mm]
\delta C_{\mu a4} &=& \partial_{\mu}\Sigma_{a4} + \omega \,\tau^b_{4a}
C_{b4\mu} - G_{\mu}\tau^c_{4a}\Sigma_{c4} - \tau_{a4}^{b}\Sigma_{\mu b}.
\label{varC4a4}
\end{eqnarray}
We assign to the gauge generators the following labels
\begin{equation}
T_{A} = \{ Z, Y, X^a, W^a \},
\label{genD4}
\end{equation}
matching the parameters $\{ \omega, \lambda, \Lambda_a, 
\Sigma_{a4} \}$ respectively.
The resulting gauge algebra is then
\begin{equation}
\begin{array}{rcl}
[Z,X^a] &=& \tau_{b4}^a X^b,\\[2mm]
[Z,W^a] &=& \tau_{b4}^a W^b.
\end{array}
\label{algebra4}
\end{equation}
The action of the gauge group is embedded in the decomposition of the 
duality group as
\begin{equation}
so(6,6) \to o(1,1)^2 + sl(5) + t_{5^{(1,0)}} + t_{\bar 5^{(1,1)}} + 
t_{\overline{10}^{(0,1)}} + t_{\bar 5^{(-1,0)}} + t_{5^{(-1,-1)}} + 
t_{10^{(0,-1)}}.
\end{equation}
The Scherk--Schwarz reduction gauges a subgroup of $sl(5) +
t_{5^{(1,0)}}$.

\section{Final comments}
\label{finalsec}

In the present paper we have derived the gauged supergravity algebras
in presence of both Scherk--Schwarz and form fluxes in the case of
M--theory (${\cal N} =8$) and type II orientifolds (${\cal N} = 4$).
In gauged supergravity, the gauge group together with its symplectic
embedding in the full duality group, determines the Yukawa couplings
and scalar potential of the theory, so these can be regarded as mass
deformations of supergravity (or massive supergravities).
In the case of Scherk--Schwarz flat groups or pure flux
compactifications, one obtains, at least at the classical level,
no--scale supergravity potentials which lift some but not all the
moduli fields.
However, it has been shown by Derendinger et
al.~\cite{Derendinger:2004jn} in a ${\cal N} =1$ example that a
combined action of Scherk--Schwarz phases and fluxes can stabilize all
the moduli and provides interesting vacua with de Sitter or Anti de
Sitter phases.
It would be interesting to carry out this analysis for the more
general ${\cal N} =4$ cases here enclosed or cases with ${\cal N} =
2,1$ supersymmetry.

The combined action of Scherk--Schwarz phases and fluxes gives a 
rather interesting structure of the gauge algebra which acts as a 
lower triangular matrix in the adjoint representation.
This is due to the nilpotent character of the axion symmetries which 
however are not inert under Scherk--Schwarz rotations.
Another important feature is that, in the presence of antisymmetric 
tensors and generic Scherk--Schwarz phases and fluxes, the conditions 
on the fluxes from the integrability conditions in $D=10,11$ do not 
guarantee the closure of the gauge Lie algebra.
Instead, it provides the closure of a more general structure, a free 
differential algebra, which also explains the non closure of the 
Jacobi identities of the Lie algebra structure discussed in 
(\ref{genalgebra}) \cite{inprep}.

Another interesting case to analyze is the enlargement of the gauged 
supergravity algebras in the case of brane form fluxes that can 
appear when a brane has some longitudinal components along the 
internal torus.
Such enlarged algebra for the Heterotic theory was analyzed by Kaloper 
and Myers \cite{Kaloper:1999yr} and it is related by S--duality to 
the D9--brane orientifold case.

We have also stated that this type of reductions can be useful to
understand more general compactification manifolds leading to
interesting 4--dimensional vacua and their relations.
In this line, it is useful to remark that it has been shown that a
certain reduction of the ${\cal N} =8$ theory, with a Scherk--Schwarz
flat group is actually identical to the type IIB ${\cal N} = 4$
${\mathbb T}^{0} \times {\mathbb T}^6$ orientifold, at least at the
level of the effective theory \cite{DAuria:2003jk}.
It would be interesting to understand if some other relation occurs
for more general compactifications on twisted tori with suitable form
fluxes turned on, along the lines of \cite{Kachru:2002sk}.
This may shed light on the duality of different compactification
manifolds.


\bigskip \bigskip

\noindent
{\bf Acknowledgments}

\medskip

We would like to thank L.~Andrianopoli, J.~P.~Derendinger, 
C.~Kounnas, M.~Lledo,
M.~Schnabl, M.~Trigiante and especially C.~Angelantonj, R.~D'Auria and F.~Zwirner
for enlightening discussions.
The work of S.F.~has been supported in part by European Community
Human Potential Program under contract MRTN-CT-2004-005104
Constituents, fundamental forces and symmetries of the universe, in
association with INFN Frascati National Laboratories and by D.O.E.
grant DE-FG03-91ER40662, Task C.



\begingroup\begin{thebibliography}{10}

\bibitem{Gukov:1999gr}
S.~Gukov, {\it Solitons, superpotentials and calibrations},  {\em Nucl. Phys.}
  {\bf B574} (2000) 169--188 [\href{http://arXiv.org/abs/hep-th/9911011}{{\tt
  hep-th/9911011}}].
S.~Gukov, C.~Vafa and E.~Witten,
Nucl.\ Phys.\ B {\bf 584} (2000) 69
[Erratum-ibid.\ B {\bf 608} (2001) 477]
[arXiv:hep-th/9906070].

\bibitem{Cardoso:2002hd}
G.~L. Cardoso, G.~Dall'Agata, D.~L\"ust, P.~Manousselis and G.~Zoupanos, {\it
  {Non-Kaehler} string backgrounds and their five torsion classes},  {\em Nucl.
  Phys.} {\bf B652} (2003) 5--34
  [\href{http://arXiv.org/abs/hep-th/0211118}{{\tt 
hep-th/0211118}}].\\
G.~L. Cardoso, G.~Curio, G.~Dall'Agata and D.~L\"ust, {\it {BPS} action and
  superpotential for heterotic string compactifications with fluxes},  {\em
  JHEP} {\bf 10} (2003) 004 [\href{http://arXiv.org/abs/hep-th/0306088}{{\tt
  hep-th/0306088}}].\\
G.~L.~Cardoso, G.~Curio, G.~Dall'Agata and D.~L\"ust,
{\it Heterotic string theory on non-Kaehler manifolds with H-flux and  gaugino
condensate,}
Fortsch.\ Phys.\  {\bf 52} (2004) 483
[arXiv:hep-th/0310021].


\bibitem{Becker:2003sh}
K.~Becker, M.~Becker, P.~S. Green, K.~Dasgupta and E.~Sharpe, {\it
  Compactifications of heterotic strings on {non-Kaehler} complex manifolds.
  {II}},  {\em Nucl. Phys.} {\bf B678} (2004) 19--100
  [\href{http://arXiv.org/abs/hep-th/0310058}{{\tt hep-th/0310058}}]. 
%
K.~Becker, M.~Becker, K.~Dasgupta and S.~Prokushkin, {\it Properties of
  heterotic vacua from superpotentials},  {\em Nucl. Phys.} {\bf B666} (2003)
  144--174 [\href{http://arXiv.org/abs/hep-th/0304001}{{\tt hep-th/0304001}}].


\bibitem{Scherk:1979zr}
J.~Scherk and J.~H.~Schwarz,
Phys.\ Lett.\ B {\bf 82}, 60 (1979).
J.~Scherk and J.~H. Schwarz, {\it How to get masses from extra dimensions},
  {\em Nucl. Phys.} {\bf B153} (1979) 61--88.

\bibitem{Derendinger:2004jn}
J.-P. Derendinger, C.~Kounnas, P.~M. Petropoulos and F.~Zwirner, {\it
  Superpotentials in iia compactifications with general fluxes},
  \href{http://arXiv.org/abs/hep-th/0411276}{{\tt hep-th/0411276}}.

\bibitem{Kaloper:1999yr}
N.~Kaloper and R.~C. Myers, {\it The O(dd) story of massive supergravity},
  {\em JHEP} {\bf 05} (1999) 010
  [\href{http://arXiv.org/abs/hep-th/9901045}{{\tt hep-th/9901045}}].

\bibitem{Kachru:2002sk}
S.~Kachru, M.~B. Schulz, P.~K. Tripathy and S.~P. Trivedi, {\it New
  supersymmetric string compactifications},  {\em JHEP} {\bf 03} (2003) 061
  [\href{http://arXiv.org/abs/hep-th/0211182}{{\tt hep-th/0211182}}].

\bibitem{Schulz:2004ub}
M.~B.~Schulz, {\it Superstring orientifolds with torsion: O5
orientifolds of torus fibrations and their massless spectra,}
Fortsch.\ Phys.\  {\bf 52} (2004) 963
[arXiv:hep-th/0406001].

\bibitem{Angelantonj:2003rq}
C.~Angelantonj, S.~Ferrara and M.~Trigiante, {\it New {D = 4} gauged
  supergravities from {N = 4} orientifolds with fluxes},  {\em JHEP} {\bf 10}
  (2003) 015 [\href{http://arXiv.org/abs/hep-th/0306185}{{\tt
  hep-th/0306185}}].

\bibitem{Angelantonj:2003up}
C.~Angelantonj, S.~Ferrara and M.~Trigiante, {\it Unusual gauged supergravities
  from type {IIA} and type {IIB} orientifolds},  {\em Phys. Lett.} {\bf B582}
  (2004) 263--269 [\href{http://arXiv.org/abs/hep-th/0310136}{{\tt
  hep-th/0310136}}].

\bibitem{Cremmer:1979uq}
E.~Cremmer, J.~Scherk and J.~H.~Schwarz,
Phys.\ Lett.\ B {\bf 84}, 83 (1979).


\bibitem{Andrianopoli:2002aq}
L.~Andrianopoli, R.~D'Auria, S.~Ferrara and M.~A. Lled\`o, {\it Duality and
  spontaneously broken supergravity in flat backgrounds},  {\em Nucl. Phys.}
  {\bf B640} (2002) 63--77 [\href{http://arXiv.org/abs/hep-th/0204145}{{\tt
  hep-th/0204145}}]. 
%
L.~Andrianopoli, R.~D'Auria, S.~Ferrara and M.~A. Lled\`o, {\it Gauging of flat
  groups in four dimensional supergravity},  {\em JHEP} {\bf 07} (2002) 010
  [\href{http://arXiv.org/abs/hep-th/0203206}{{\tt hep-th/0203206}}].

\bibitem{Louis:2002ny}
J.~Louis and A.~Micu, {\it Type {II} theories compactified on {Calabi-Yau}
  threefolds in the presence of background fluxes},  {\em Nucl. Phys.} {\bf
  B635} (2002) 395--431
  [\href{http://arXiv.org/abs/http://arXiv.org/abs/hep-th/0202168}{{\tt
  hep-th/0202168}}].

\bibitem{DallAgata:2003yr}
G.~Dall'Agata, R.~D'Auria, L.~Sommovigo and S.~Vaul\`a, {\it {D = 4, N = 2}
  gauged supergravity in the presence of tensor multiplets},
  \href{http://arXiv.org/abs/hep-th/0312210}{{\tt hep-th/0312210}}.

\bibitem{DAuria:2004yi}
R.~D'Auria, L.~Sommovigo and S.~Vaul\`a, {\it {N = 2} supergravity {Lagrangian}
  coupled to tensor multiplets with electric and magnetic fluxes},  {\em JHEP}
  {\bf 11} (2004) 028 [\href{http://arXiv.org/abs/hep-th/0409097}{{\tt
  hep-th/0409097}}].

\bibitem{Louis:2004xi}
J.~Louis and W.~Schulgin, {\it Massive tensor multiplets in N = 1
  supersymmetry},  \href{http://arXiv.org/abs/hep-th/0410149}{{\tt
  hep-th/0410149}}.

\bibitem{DAuria:2004tr}
R.~D'Auria, S.~Ferrara, M.~Trigiante and S.~Vaul\`a, {\it Gauging the heisenberg
  algebra of special quaternionic manifolds},
  \href{http://arXiv.org/abs/hep-th/0410290}{{\tt hep-th/0410290}}.

\bibitem{Sommovigo:2005fk}
L.~Sommovigo, {\it Poincar\'e dual of D = 4 N = 2 supergravity with tensor
  multiplets},  \href{http://arXiv.org/abs/hep-th/0501048}{{\tt
  hep-th/0501048}}.

\bibitem{inprep}
G.~Dall'Agata, R.~D'Auria and S.~Ferrara, to appear.
  
\bibitem{Solva}
L.~Andrianopoli, R.~D'Auria, S.~Ferrara, P.~Fre and M.~Trigiante,
{\it R-R scalars, U-duality and solvable Lie algebras,}
Nucl.\ Phys.\ B {\bf 496}, 617 (1997)
[arXiv:hep-th/9611014].
L.~Andrianopoli, R.~D'Auria, S.~Ferrara, P.~Fre, R.~Minasian and M.~Trigiante,
{\it Solvable Lie algebras in type IIA, type IIB and M theories,}
Nucl.\ Phys.\ B {\bf 493}, 249 (1997)
[arXiv:hep-th/9612202].


\bibitem{Andrianopoli:2004sv}
L.~Andrianopoli, S.~Ferrara and M.~A. Lled\`o, {\it Axion gauge symmetries and
  generalized {Chern-Simons} terms in N = 1 supersymmetric theories},  {\em
  JHEP} {\bf 04} (2004) 005 [\href{http://arXiv.org/abs/hep-th/0402142}{{\tt
  hep-th/0402142}}].

\bibitem{Cremmer:1979up}
E.~Cremmer and B.~Julia, {\it The $SO(8)$ Supergravity}, Nucl.\ Phys.\ B
{\bf 159}, 141 (1979).

\bibitem{Frey:2002hf}
A.~R.
Frey and J.~Polchinski, {\it {$N = 3$} warped compactifications}, {\em
Phys.
Rev.} {\bf D65} (2002) 126009
[\href{http://arXiv.org/abs/http://arXiv.org/abs/hep-th/0201029}{{\tt
hep-th/0201029}}].

\bibitem{DAuria:2003jk}
R.~D'Auria, S.~Ferrara, F.~Gargiulo, M.~Trigiante and S.~Vaul\`a, {\it N
= 4 supergravity Lagrangian for type IIB on $T^6/Z_2$ in presence of
fluxes and D3-branes,} JHEP {\bf 0306} (2003) 045
{\tt [hep-th/0303049]}.

\bibitem{Andrianopoli:2005jv}
L.~Andrianopoli, M.~A.~Lledo' and M.~Trigiante, {\it The
Scherk-Schwarz mechanism as a flux compactification with internal
torsion}, {\tt hep-th/0502083}.


\end{thebibliography}\endgroup
\providecommand{\href}[2]{#2}

\end{document}